\documentclass[twoside,a4paper,11pt]{sea22}
\usepackage{graphicx}
\usepackage{hyperref}
\usepackage{media9}
\usepackage{amssymb}
\begin{document}
\pagenumbering{arabic}
\pagestyle{myheadings}
\thispagestyle{empty}
{\flushleft\includegraphics[width=\textwidth,bb=58 650 590 680]{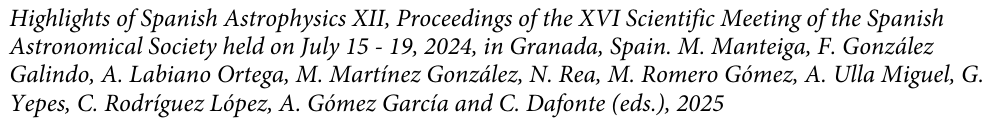}}
\vspace*{0.2cm}
\begin{flushleft}
{\bf {\LARGE
%
%%% TITLE of the paper. 
%%% TITLE of the paper. 
Growth of Diffuse Intragroup Light in Simulated Galaxy Groups
%
% Do not delete next few lines
}\\
\vspace*{1cm}
%
%%% Include here the LIST OF AUTHORS.
%%% Include here the LIST OF AUTHORS.
%%% Note that the last author has to be preceeded by an AND.
Bilata-Woldeyes, B.$^{1}$,
Perea, J.D.$^{1}$ and 
Solanes, J.M.$^{2}$
%
% Do not delete next few lines
}\\
\vspace*{0.5cm}
%
%%% AFFILIATIONS LIST.
%%% and the AFFILIATIONS LIST. Note that one affiliation per line.
%%% Add as many affiliations as necessary. 
$^{1}$
Instituto de Astrof\'\i sica de Andaluc\'\i a (IAA–CSIC), Glorieta de la Astronom\'\i a, s/n, E-18008 Granada, Spain\\
$^{2}$
Departament de F\'\i sica Qu\`antica i Astrof\'\i sica and Institut de Ci\`encies del Cosmos (ICCUB), Universitat de Barcelona. C.\ Mart\'{\i}  i Franqu\`es, 1, E-08028 Barcelona, Spain
%
% Do not delete next few lines
\end{flushleft}
%
% Headings
\markboth{
%%% Type the SHORT version of the paper title.
%%% Type the SHORT version of the paper title.
%Short version of the paper title
Growth of Diffuse Intergalactic Light
}{ % Do not delete
%
%%%  First Author \& Second Author   OR   First-author et al. 
%%%  First Author \& Second Author   OR   First-author et al. if the author list 
%%% contains three or more authors.
%Author 1 \& Author 2  OR  Author 1 et al. if three or more authors
Bilata-Woldeyes et al.
% 
% Do not delete next few lines
}
\thispagestyle{empty}
\vspace*{0.4cm}
\begin{minipage}[l]{0.09\textwidth}
\ 
\end{minipage}
\begin{minipage}[r]{0.9\textwidth}
\vspace{1cm}
\section*{Abstract}{\small
%
% ABSTRACT ABSTRACT ABSTRACT
% ABSTRACT ABSTRACT ABSTRACT
%%% Type the ABSTRACT of your oral contribution or poster
The diffuse intragroup light (IGL) is a pervasive feature of galaxy groups consisting of an extended low-surface-brightness component that permeates the intergalactic medium of these galaxy associations. It is primarily formed by stars that are separated from their host galaxies and now drift freely, unbound to any particular galaxy. We used controlled numerical simulations to investigate the formation and evolution of IGL in galaxy groups during the pre-virialization phase. Our study reveals that the emergence of this diffuse luminous component typically begins to form in significant amounts around the turnaround epoch, increasing steadily thereafter. We analyzed the correlation between the mass and fraction of IGL and other group properties, finding a sublinear relationship between the IGL mass and the brightest group galaxy, suggesting intertwined formation histories but with potentially differing growth rates and distinct driving mechanisms. Additionally, we observed a negative correlation between the IGL fraction and group’s velocity dispersion, indicating that a lower velocity dispersion may enhance IGL formation through increased gravitational interaction effectiveness. Furthermore, our comparative analysis of the density profiles of IGL and total system mass revealed significant similarities, suggesting that the intragroup light serves as a reliable tracer of the gravitational potential of host groups, even when these galaxy aggregations are far from dynamic equilibrium.
%
% Do not delete next few lines
\normalsize}
\end{minipage}
%
%
%%% BODY of the paper
%%% BODY of the paper
%
\section{Introduction \label{intro}}
Deep surface photometry reveals the presence in a good number of galaxy groups, as well as in their larger counterparts, galaxy clusters, of an extended, diffuse luminous component that fills the space between galaxies. This intragroup light (IGL) is formed by stars that have been incorporated into the internal medium of these galaxy associations after being separated from their host galaxies \cite{2006ApJ...648..936R, 2019A&A...622A.183J}, or created in situ during the disruptive gravitational interactions that member galaxies experience during the formation stage of these systems \cite{2016MNRAS.461..321S}. Often found concentrated around the central, brightest group galaxy (BGG) \cite{2022NatAs...6..308M}, the IGL is a valuable source of information regarding the group's built-up and evolutionary stage \cite{2006ApJ...648..936R}. Additionally, several investigations suggest that this luminous component may also trace the dark matter distribution within these galaxy systems independently of their dynamical state \cite{2019MNRAS.482.2838M, 2020ApJ...901..128C, 2022ApJS..261...28Y}.

In this work, we analyzed the influence of various group parameters on the formation of IGL and its correlation. Additionally, we explored the potential of using the IGL as a luminous tracer for the overall gravitational potential of the host galaxy aggregations, particularly in cases where these systems are not in dynamic equilibrium.  

\section{Numerical simulations}
We ran a series of 100 galaxy group simulations that trace the early formation stages of these systems, focusing on the production of the ex-situ IGL that occurs before they reach full dynamical relaxation. The simulated groups were created as nearly uniform isolated spherical overdensities at z = 3 that first expand linearly, then turnaround, and finally undergo a completely non linear collapse at z = 0, which marks the end of the simulation (see \cite{2016MNRAS.461..321S, 2016MNRAS.461..344P}). 

Among the various methods proposed in the literature to quantify the IGL fraction, defined by the ratio of the IGL to the total light, the application of an upper limit on the surface brightness (SB) to separate galactic from intergalactic starlight is by far the most extensively used technique \cite{2018ApJ...862...95K, 2022NatAs...6..308M, 2022ApJ...925..103C}. Simulations have demonstrated the effectiveness of this approach for SB values fainter than $\mu_V \gtrsim 26.0$--$26.5$ mag arcsec$^{-2}$, roughly corresponding to the level of the Holmberg radius, concurrently revealing that most of the IGL fraction has been produced in a relatively recent cosmic epoch 
($z\lesssim 1$; e.g. \cite{2011ApJ...732...48R, 2014MNRAS.437.3787C, 2019A&A...622A.183J, 2021MNRAS.502.2419F, 2022NatAs...6..308M, 2022ApJ...925..103C}). Following \cite{2005ApJ...631L..41M}, \cite{2011ApJ...732...48R}, and \cite{2014MNRAS.437..816C}, among others, we applied a projected stellar density threshold equivalent to a SB of $26.5$ mag~arcsec$^{-2}$ in the $V$ band. 

\section{Results}
\subsection{Time evolution of the IGL fraction} 
\begin{figure}[!h]
\center 
\includegraphics[width=7.5cm,height=4cm]{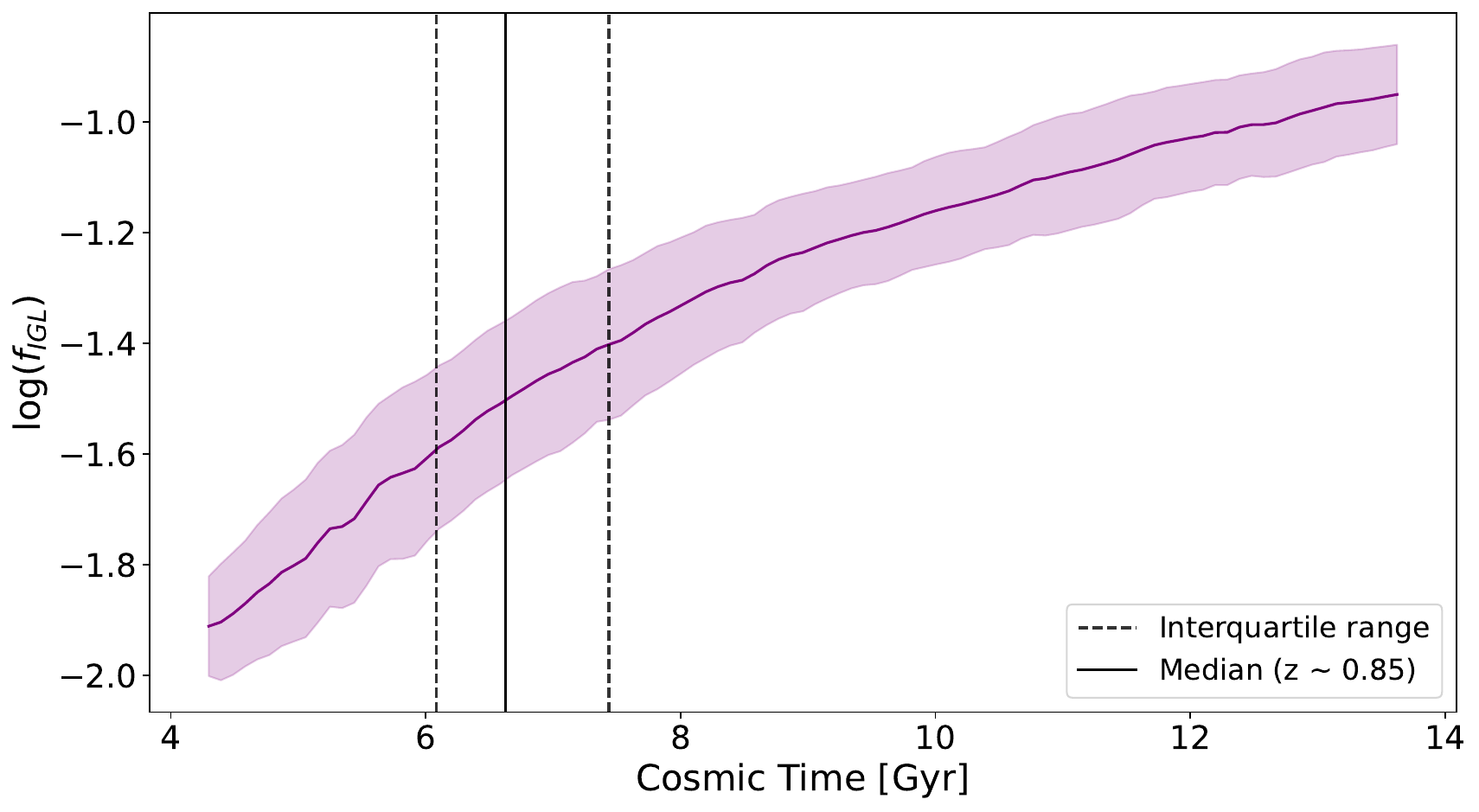} 
\caption{\label{fig1} \small{Evolution  of the IGL fraction over cosmic time. The purple band represents the interquartile range of the temporal variation in the distribution of $\log(f_{\rm{IGL}})$ with the central solid line indicating the evolution of the median value. The vertical solid and dashed lines mark, respectively, the $Q_2$, $Q_1$, and $Q_3$ quartiles of the distribution of turnaround times, when the groups reach their maximum expansion.}}
\end{figure}

As shown in Fig.~\ref{fig1}, $f_{\rm IGL}$ grows steadily with cosmic time. Although the relative growth of diffuse light is faster and roughly linear at early times, the majority of this component is produced after the groups reach their turnaround epoch (at a median $z\sim 0.85$ in our simulations), consistent with previous studies stating that the IGL is mainly produced at $z < 1$ \cite{2007MNRAS.377....2M, 2011ApJ...732...48R, 2015MNRAS.449.2353B, 2016MNRAS.461..321S}. Beyond this epoch, the ongoing hierarchical collapse of these systems shortens intergalactic distances, increasing the frequency and intensity of gravitational interactions and enhancing the efficiency of stellar stripping experienced by their member galaxies (e.g. \cite{2018MNRAS.474.3009D, 2019A&A...622A.183J}). The quartiles of the distribution of the total IGL mass generated in our groups at $z = 0$ are $(Q_1,Q_2,Q_3)=(7.0,9.8,13)\,\times 10^{10}\,\mbox{M}_{\odot}$, while the corresponding IGL fraction $(f_{\rm{IGL}})$ ranges between $6.5$--$18\%$, with a median value of $11.4\%$.
%-------------------------
%----------------------------

\subsection{BGG-based classification of groups}
\begin{figure}[!h]
\center 
\includegraphics[width=8.5cm,height=5cm]{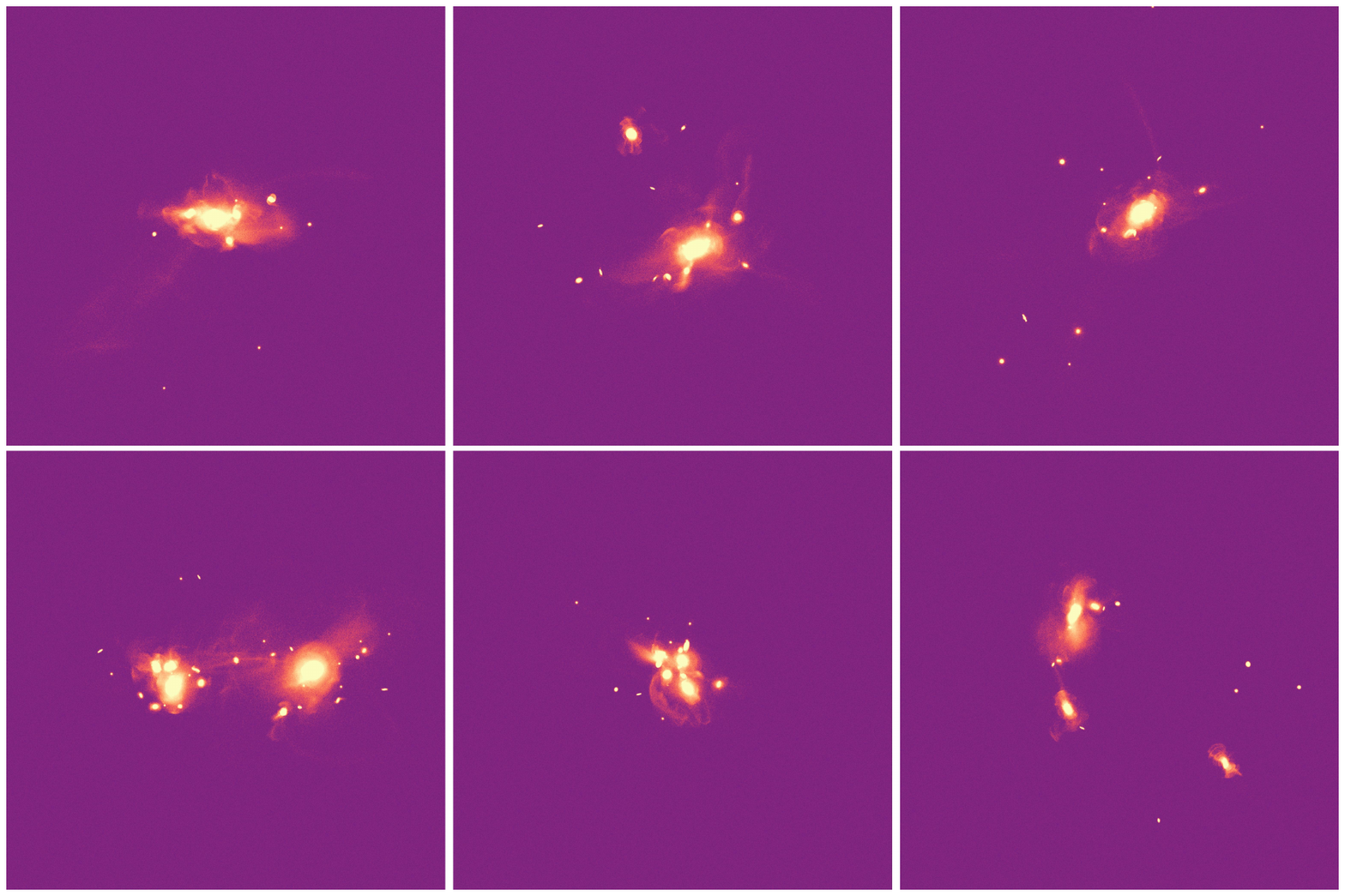} 
\caption{\label{fig2} \small{Examples of group images at z = 0. Upper panels: groups with a clearly dominant galaxy or BGG. Lower panels: groups with two dominant galaxies (left) or lacking a truly dominant BGG (center and right). Note that in all cases, the IGL is mainly distributed around the largest galaxies.}}
\end{figure}

Visual inspection of the group images at $z = 0$ reveal that each particular realization of the initial conditions of our group model results in a wide variety of assembly paths (see, Fig.~\ref{fig2}). Therefore, the groups are classified based on the different types of first-ranked galaxies\footnote{First-ranked galaxies ($m_1$) are the most massive galaxies in the group in terms of stellar mass. Those with a magnitude gap with respect to the second most massive object ($m_2$) above a certain threshold are true
BGGs.} hosted by the groups. These parameters compare the luminous masses of the three largest group members, expressed in the familiar form of magnitude differences between the first- and second-ranked galaxy, $\Delta{\cal{M}}_{\rm{2-1}}$, and between the second- and third-ranked galaxy $\Delta{\cal{M}}_{\rm{3-2}}$. We identified a subset of groups, termed single-BGG, which host a clearly dominant galaxy separated from the rest by at least $0.75$ magnitudes at $z=0$. This class comprises $38\%$ of the group images at this redshift. Additionally, $16\%$ of our systems are classified as double-BGG, characterized by $\Delta{\cal{M}}_{2-1} < 0.5$ and $\Delta{\cal{M}}_{3-2} \ge 0.75$ indicating the presence of a pair of bright galaxies that stand out from the rest of group members. The remaining $46\%$ of the $z=0$ images correspond to groups categorized as non-BGG, as they have both $\Delta{\cal{M}}_{2-1}$ and $\Delta{\cal{M}}_{3-2} < 0.75$. These latter systems, often the most massive in our simulations, typically feature multiple ongoing mergers but lack a single dominant galaxy.
%-----------------------------
%--------------------------------

\subsection{IGL correlation with other group properties at $z=0$}
\begin{figure}[!h]
\center 
\includegraphics[width=9.5cm,height=4.5cm]{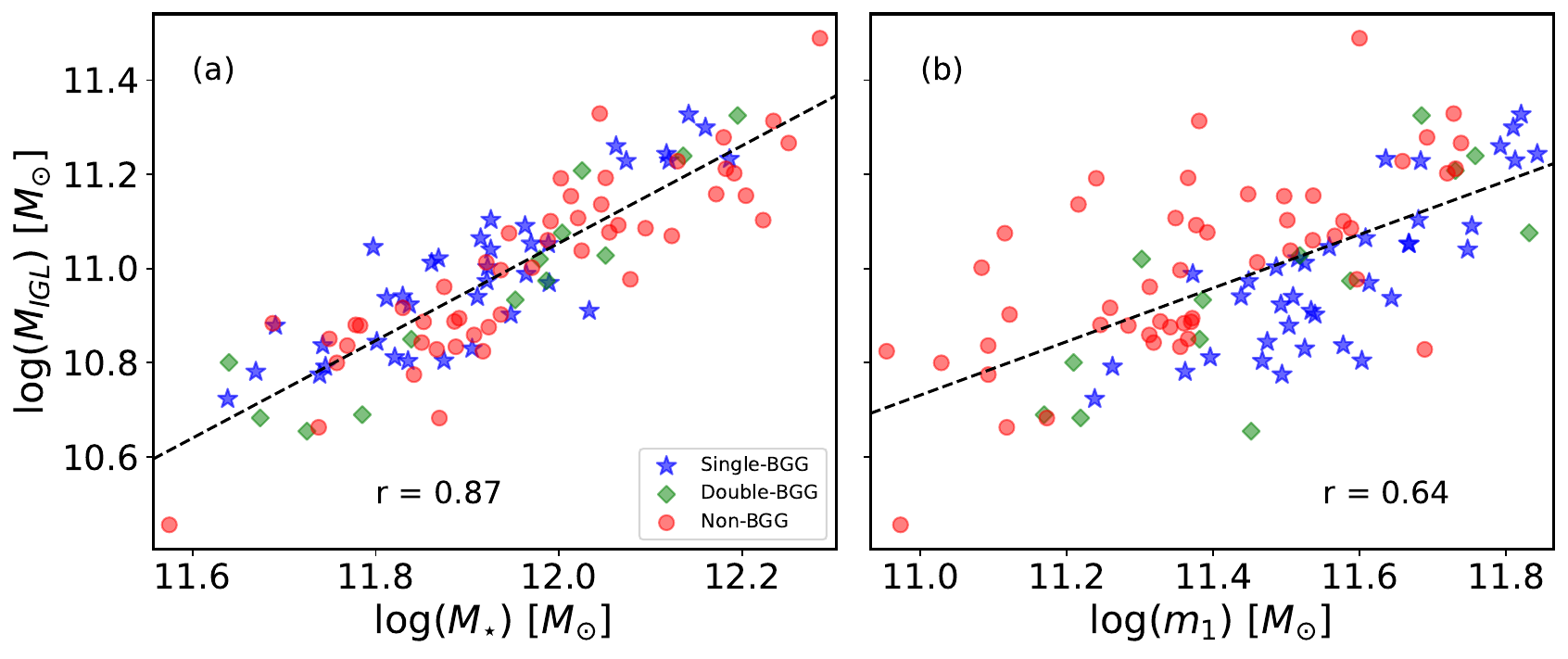}
\includegraphics[width=5.8cm,height=4.5cm]{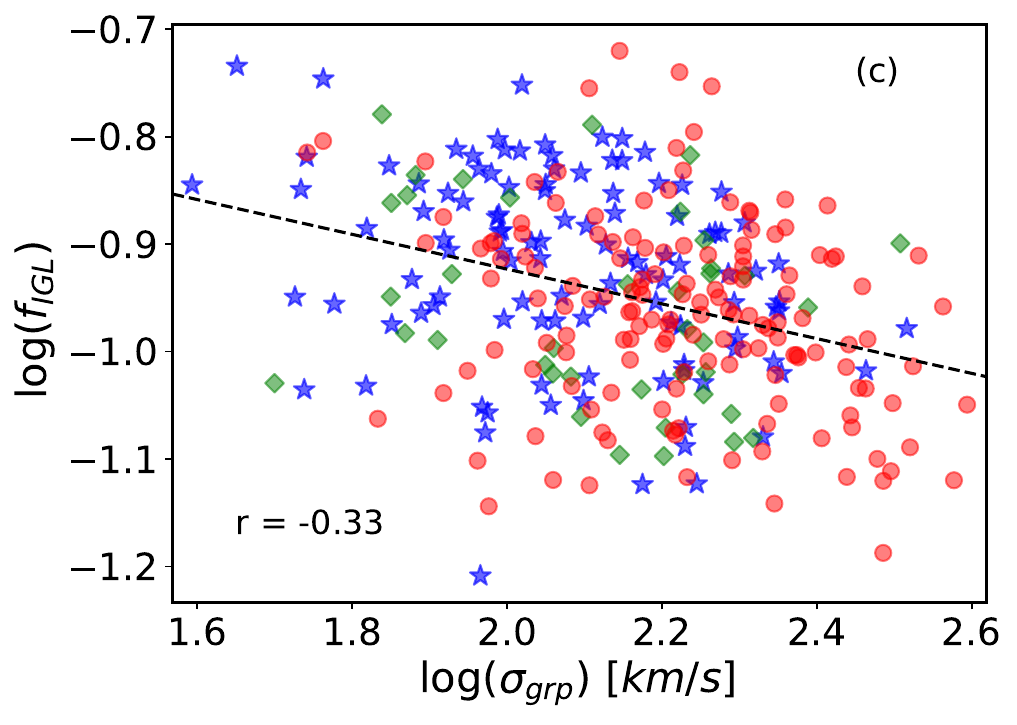}
\caption{\label{fig3} \small{Scatter plots of $M_{\mathrm{IGL}}$ against, (a) total stellar mass ($M_\star$) and (b) mass of the first-ranked galaxy ($m_{1}$) and, panel (c) shows the $f_{IGL}$ correlation with the groups' velocity dispersion ($\sigma_{grp}$), all measured at $z = 0$. Symbols for data points correspond to the single-BGG (blue-star), double-BGG (green-diamond), and non-BGG (red-circle) classifications. The values of Pearson's correlation coefficient $r$ are included in the plots.}}
\end{figure}

A strongest positive correlations is observed with the mass of the IGL and total stellar mass of the group with $r=0.87$ (see Fig.~\ref{fig3}a) suggesting, the more stellar component in the group the higher the mass of the IGL. In panel b, it is shown that the growth of $M_{IGL}$ with $m_{\rm{1}}$ follows a sublinear power-law of exponent $0.57$. While looking at the correlation of the $M_{\rm{IGL}}$ with the mass of the first-ranked galaxy for those groups with single and double BGGs, a somewhat stronger positive correlation is observed, but nonetheless still sublinear with a power-law exponent of $0.81$. In our simulations, the groups with more massive BGGs tend to produce a larger IGL fraction.

Fig.~\ref{fig3}c shows the correlation between the $f_{\rm{IGL}}$ with the velocity dispersion of the groups $\sigma_{\rm{grp}}$ measured along their three principal axes. Despite of its very high $p$-value\footnote{The 
$p$-value indicates the likelihood of observing a non-zero linear correlation coefficient in our sample data under the assumption that the null hypothesis is true.} ($p \sim 1$), this correlation is rather weak ($r = -0.33$), the negative slope suggesting that slower gravitational interactions among galaxies within groups are most effective at increasing the fraction of IGL, as they are more efficient at stripping material from galaxy discs (see, e.g. \cite{2003astro.ph..5512M, 2016ApJS..225...23S}). 
%-----------------------------
%--------------------------------

\subsection{Density profile of the IGL and the total system mass}
\begin{figure}[!h]
\center 
\includegraphics[width=6cm,height=4cm]{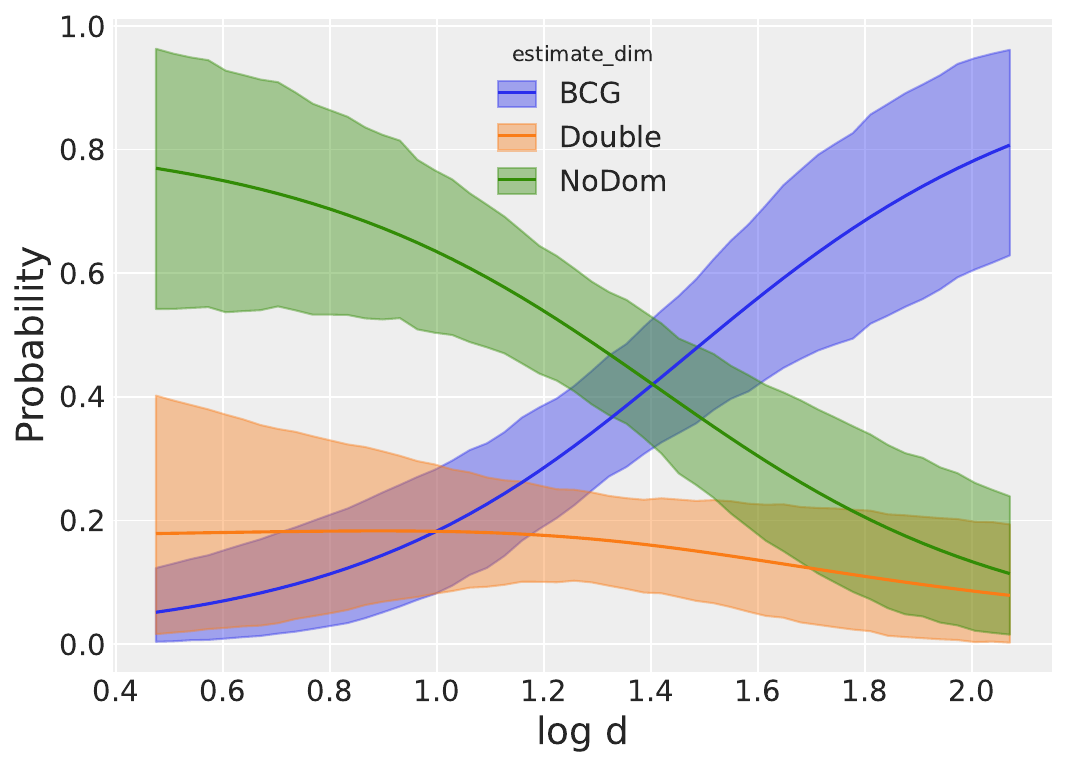}
\includegraphics[width=8cm,height=4cm]{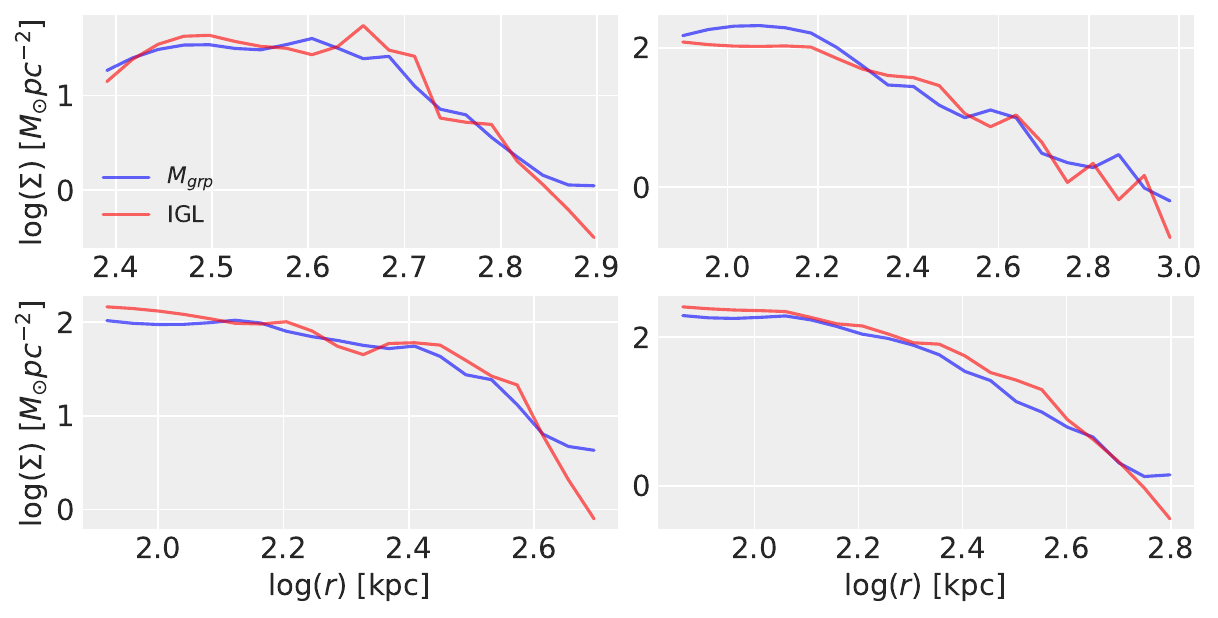}
\caption{\label{fig4} \small{Left panel, probability that the profile distance ($d$) takes on a certain value according to the BGG class of the group. The blue, orange and green bands encompass interquartile ranges of the probability for single-BGG, double-BGG, and non-BGG groups, respectively. Right panels: four examples of the IGL (red) and total system mass (blue) projected radial density profiles in two single-BGG (top) and two non-BGG (bottom) groups with $d < 20\%$.}}
\end{figure}

In recent works \cite{2019MNRAS.482.2838M, 2020ApJ...901..128C, 2022ApJS..261...28Y}, it has been claimed that the IGL can be used as a tracer for the distribution of the total mass of a system. To verify this, we have generated and compared the shapes of the radial distributions of the azimuthally averaged surface density for the IGL and total mass. The discrepancy between the density profiles for individual groups has been quantified through a metric which provides a direct quantification of the 'fractional distance’ between the shapes of the
datasets that are being compared:
%------------------------------
\begin{equation}\label{d}
d(X,Y) = \mbox{med}\left(\left|\frac{\log Y(R) - \log X(R)}{\log X(R)}\right|\right)\;,   
\end{equation}
%--------------------------
where the $X(R)$ and $Y(R)$ subsets represent the binned radial surface density profiles of the total and IGL mass, respectively, and $R$ the projected group-centric bin radii. The median (Q2) of the distribution of these fractional distances for our simulated groups is $21\%$, with $Q1 = 12.5\%$ and $Q3 = 31.9\%$. Groups with non-BGG class tend to show lower profile distances than groups with single-BGG. For their part, double-BGG systems show a flat trend in $d$, with a mild tendency to favor lower profile distances (see left panel of \ref{fig4}).
%-----------------------------
%--------------------------------

\section{Conclusions} 
\begin{itemize}
\item The IGL begins to form in substantial amounts around turnaround, $z \sim 0.85$ and keeps growing steadily with cosmic time until the end of simulations at $z = 0$.
\item The mass of IGL increases almost linearly ($r = 0.87$) with the total stellar mass of the group.
\item While the growth of the IGL is linked to the development of the first-ranked galaxy, its formation is not necessarily tied to the presence of a dominant BGG, as considerable amounts of IGL are observed in some non-BGG groups.
\item Slower gravitational interactions among galaxies within groups are most effective at increasing the IGL fraction due to their higher efficiency in stripping material from galaxy disks.
\item Our simulations corroborate previous observational findings showing that the IGL serves as a reliable tracer of the gravitational potential of galaxy aggregations, particularly those with high total mass, even when these systems are far from dynamic equilibrium.
\end{itemize}

%
%
% Do not delete the next line
\small  % Do not delete
%
%%% Comment the following line if you do not have acknowledgments.
\section*{Acknowledgments}   % Do not delete if you declare acknowledgments
%
%%% ACKNOWLEDGMENTS
%%% ACKNOWLEDGMENTS
We acknowledge financial support from the Spanish state agency MCIN/AEI/10.13039/501100011033 and by 'ERDF A way of making Europe' funds through research grants PID2022-140871NB-C21 and PID2022-140871NB-C22. MCIN/AEI/10.13039/501100011033 has also provided additional support through the Centre of Excellence Severo Ochoa's award for the Instituto de Astrof\'\i sica de Andalucía under contract CEX2021-001131-S and the Centre of Excellence Mar\'\i a de Maeztu's award for the Institut de Ci\`encies del Cosmos at the Universitat de Barcelona under contract CEX2019–000918–M. BBW acknowledges financial support by the PRE2020-093715 grant from the SEV–2017–0709 project. 

%
% Do not delete the next few lines

%
\end{document}